\documentclass[journal]{IEEEtran}
\usepackage{amsmath,amsfonts}
\usepackage{algorithmic}
\usepackage{algorithm}
\usepackage{array}
\usepackage[caption=false,font=normalsize,labelfont=sf,textfont=sf]{subfig}
\usepackage{textcomp}
\usepackage{stfloats}
\usepackage{url}
\usepackage{verbatim}
\usepackage{graphicx}
\usepackage{cite}
\usepackage{times}
\usepackage{epsfig}
\usepackage{graphicx}
\usepackage{amsmath}
\usepackage{amssymb}
\usepackage{multirow}
\usepackage{pifont}
\usepackage{arydshln}
\usepackage{booktabs}
\usepackage{bbding}
\usepackage{caption}
\PassOptionsToPackage{dvipsnames,svgnames}{xcolor}
\usepackage{xcolor}
\usepackage[percent]{overpic}
\usepackage{ragged2e}

\usepackage[accsupp]{axessibility} 

\usepackage{color}
\usepackage{amsmath}

\usepackage{colortbl}  
\usepackage{xcolor}
\usepackage{array}

\newcommand{\eg}{e.g.}
\newcommand{\ie}{i.e.}
\newcommand{\etc}{etc}
\newcommand{\q}[1]{\textit{``#1''}}

\definecolor{hollywoodcerise}{rgb}{0.96, 0.0, 0.63}
\definecolor{lasallegreen}{rgb}{0.03, 0.47, 0.19}
\definecolor{hanpurple}{rgb}{0.32, 0.09, 0.98}
\definecolor{green(pigment)}{rgb}{0.0, 0.65, 0.31}
\usepackage{hyperref}

\begin{document}


\title{Learning High-Quality Navigation and Zooming on Omnidirectional Images in Virtual Reality}

\author{Zidong Cao$^{1}$, Zhan Wang$^{1}$, Yexin Liu$^{1}$, Yan-Pei Cao$^{3}$, Ying Shan$^{3}$, Wei Zeng$^{1,2}$ and Lin Wang$^{1,2*}$
\thanks{$^{1}$HKUST(GZ)~~$^{2}$HKUST~~$^{3}$ARC Lab, Tencent PCG.}
\thanks{$^*$Corresponding author.}}


\markboth{Journal of \LaTeX\ Class Files,~Vol.~14, No.~8, August~2021}%
{Shell \MakeLowercase{\textit{et al.}}: A Sample Article Using IEEEtran.cls for IEEE Journals}


\maketitle

\begin{abstract}
Viewing omnidirectional images (ODIs) in virtual reality (VR) represents a novel form of media that provides immersive experiences for users to navigate and interact with digital content. Nonetheless, this sense of immersion can be greatly compromised by a blur effect that masks details and hampers the user's ability to engage with objects of interest. In this paper, we present a novel system, called \textit{\textbf{OmniVR}}, designed to enhance visual clarity during VR navigation. Our system enables users to effortlessly locate and zoom in on the objects of interest in VR. It captures user commands for navigation and zoom, converting these inputs into parameters for the M\"obius transformation matrix. Leveraging these parameters, the ODI is refined using a learning-based algorithm. The resultant ODI is presented within the VR media, effectively reducing blur and increasing user engagement. To verify the effectiveness of our system, we first evaluate our algorithm with state-of-the-art methods on public datasets, which achieves the best performance. Furthermore, we undertake a comprehensive user study to evaluate viewer experiences across diverse scenarios and to gather their qualitative feedback from multiple perspectives. The outcomes reveal that our system enhances user engagement by improving the viewers' recognition, reducing discomfort, and improving the overall immersive experience. Our system makes the navigation and zoom more user-friendly. For more details, \eg, demos, please refer to the project page \url{http://vlislab22.github.io/OmniVR/}.

\end{abstract}



\section{Introduction}

\begin{figure*}[t]
    \centering
    \includegraphics[width=0.9\linewidth]{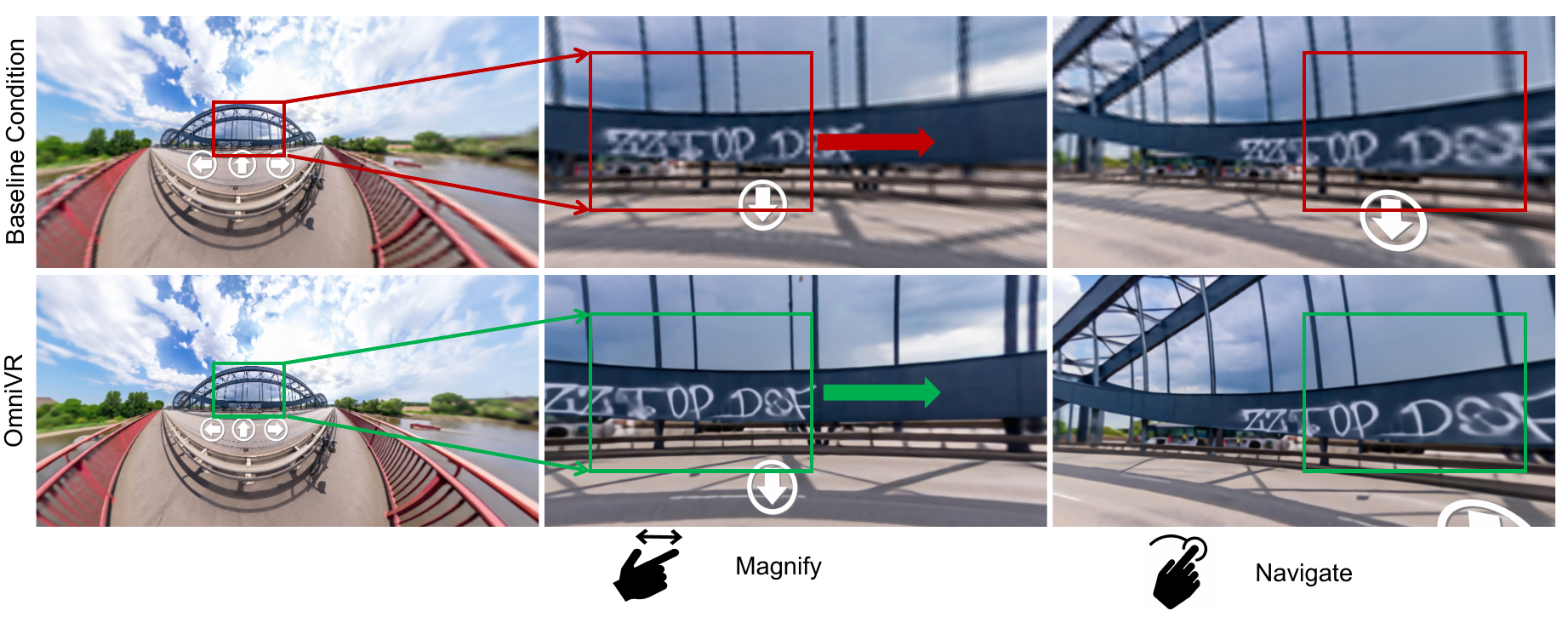}
\caption{\textbf{Comparing the VR experience under the baseline condition and our OmniVR}. Users can freely navigate and zoom in/out to see the object of interest. With our proposed algorithm, the objects can be refined with clear textural details, thus enhancing the engagement and immersive experience.}
\label{fig:teaser}
\end{figure*}


\IEEEPARstart{O}{m}nidirectional images (ODIs), also called $360^{\circ} $ images, have increasingly attracted interest for their capability to capture extensive content within a single frame. The recent surge in integrating such visual content into virtual reality (VR) environments is noteworthy~\cite{vermast2023introducing, luo2023masked360, dasari2023scaling}. This integration represents a novel form of media that allows users to navigate freely in any direction, offering an immersive and interactive experience akin to being in a real environment~\cite{zhang2022realvr,lou2019realistic,szabo2022cnn}. This immersive media has led to various applications, including but not limited to virtual tours~\cite{verma2022past, mohammad2009development}, real estate showcases~\cite{azmi2022smarter}, educational tools~\cite{singh2022metaverse}, and remote meeting solutions~\cite{dasari2023scaling}. Moreover, the rapid advancement and wide availability of consumer-level VR devices~\cite{chang2017panning} have made these experiences increasingly accessible to a broader audience. In this context, viewers transform from mere observers to active participants, who can navigate and zoom into objects of interest, thereby making significant progress in media consumption and user experience~\cite{chang2017panning}.

The ODIs are usually stored with the equirectangular projection (ERP) type and displayed in VR with perspective projection. A critical issue with ODIs is their relatively low angular resolution~\cite{yoon2022spheresr}, which results in local regions appearing blurry. This blurriness intensifies when the images are zoomed in and navigated (see Fig.~\ref{fig:teaser} (top)), potentially compromising the immersive experience by reducing local and fine details~\cite{kwon2016compensation}. Consequently, this not only impedes user engagement with objects of interest but also detracts from the overall immersive experience, potentially causing mental and physical discomfort~\cite{o2013visual}. Several methods have been explored to mitigate this discomfort in VR navigation, including the incorporation of spatial blur~\cite{hoffman2008vergence}, defocus blur~\cite{ang2020gingervr}, depth-of-field blur~\cite{hussain2021mitigating}, and foveated rendering~\cite{meng2018kernel}. However, these methods fall short of enhancing the clear textural details of objects. Despite the freedom to navigate and zoom in/out on objects, the enlarged objects remain blurry.

\begin{figure*}[t]
    \centering
    \includegraphics[width=0.98\textwidth]{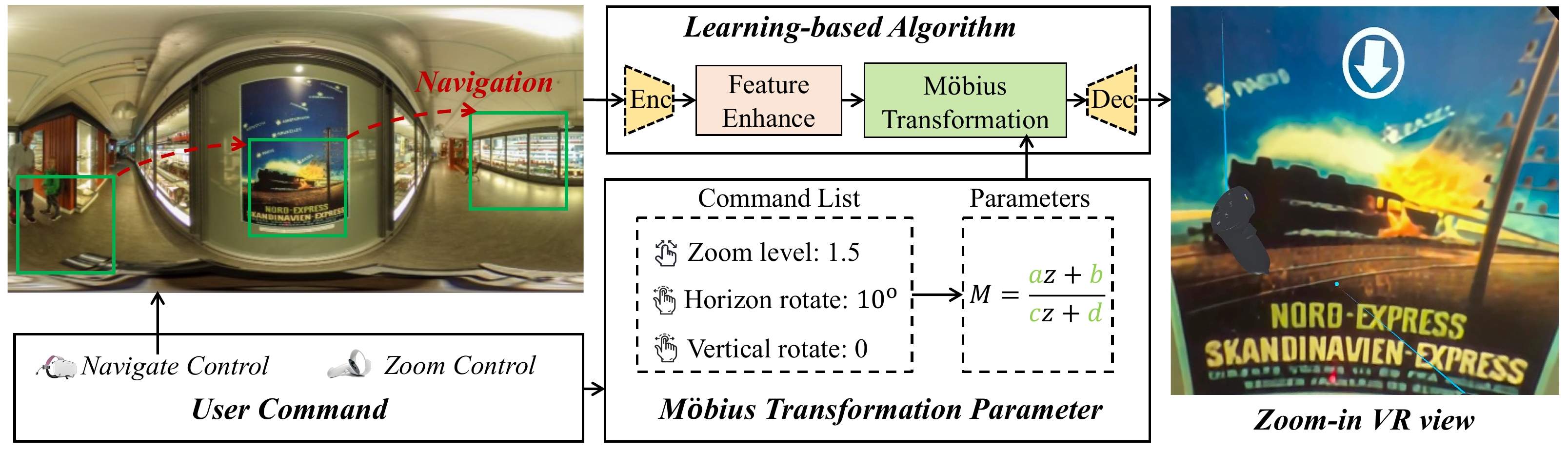}
\caption{\textbf{System overview.} Our system collects the user commands for navigation and zooming in/out, which are converted to the parameters of the M\"obius transformation matrix. The parameters together with the ODI are processed with a learning-based algorithm to generate high-quality transformed ODIs, which can be displayed with the perspective format in VR.}
\label{fig:system}
\end{figure*}

In this paper, we introduce a novel system, dubbed \textit{OmniVR}, aiming to make viewers navigate and zoom in/out in the VR space effortlessly, while simultaneously enhancing the visual quality to recover clear local details, as illustrated in Fig.~\ref{fig:teaser} (bottom). As shown in Fig.~\ref{fig:system}, the viewer first views the original ODI displayed in a VR headset. With \textit{OmniVR}, the viewer is free to navigate and find some objects of interest. Then, the viewer can use the headset and controller to give commands about rotation and zoom in/out. Our system captures these user commands and converts these commands into parameters for the M\"obius transformation matrix (Sec.~\ref{sec: system input}). Leveraging these parameters, we propose a learning-based algorithm, which is built based on our conference work OmniZoomer~\cite{cao2023omnizoomer}, to achieve high-quality ODIs after transformation with two key techniques. First, OmniVR integrates the M\"obius transformation into the network, enabling free navigation and zoom within ODIs. By learning transformed feature maps in various conditions, the network is enhanced to handle the increasing curves caused by navigation and zoom, thus alleviating the blurry effect (Sec.~\ref{method}). Secondly, we propose enhancing the feature maps to high-resolution (HR) space before the transformation. The HR feature maps contain more fine-grained textural details, which could compensate for the lack of pixels for describing curves (Sec.~\ref{sec:upsample}). After obtaining the HR feature maps, we also propose a spatial index generation module (Sec.~\ref{sec:grid}) and a spherical resampling module (Sec.~\ref{sec:Slerp}) to accomplish the feature transformation process. Finally, these feature maps are processed with a decoder to output the high-quality transformed ODI in ERP format. The ERP output is then transformed to the perspective format to be displayed in VR,
effectively reducing blur and increasing user engagement.

For supervised learning, we create a dataset based on the ODI-SR dataset~\cite{deng2021lau}, called the ODIM dataset, including transformed ODIs under various M\"obius transformations. We evaluate the effectiveness of OmniVR on the ODIM dataset under various M\"obius transformations and up-sampling factors. Furthermore, we report the results of a user study for the VR experience to evaluate the effectiveness of our proposed system in quantitative and qualitative ways. Quantitatively, we record accuracy, response time, and confidence level in a series of scenarios and questions. Qualitatively, we conduct interviews about the subjects' feelings, such as mental and physical costs, and immersive experience, \etc. The results demonstrate that: 1) \textit{OmniVR} is beneficial for participants to improve the recognition and understanding of the scenarios. 2) \textit{OmniVR} can reduce the discomfort of participants. 3) \textit{OmniVR} can significantly improve the immersive experience, and make navigation and zoom in the VR media more user-friendly.

The main contributions of this paper can be summarized as follows: (\textbf{I}) We propose a novel system \textit{OmniVR} to enhance the visual clarity during VR navigation; (\textbf{II}) We propose a learning-based algorithm to enhance the ODI quality controlled by user commands; (\textbf{III}) We establish the ODIM dataset for supervised training. Compared with existing methods, OmniVR achieves state-of-the-art performance under various user commands and up-sampling factors; (\textbf{IV}) We conduct a user study, demonstrating the effectiveness of our system in both quantitative and qualitative ways.

\section{Related Works}


\noindent \textbf{Navigation in VR.} VR is an emerging media that provides more immersive and interactive experiences compared with traditional media~\cite{zhang2022realvr,lou2019realistic,szabo2022cnn}. Based on this immersive experience, there are already various applications, spanning virtual tourism~\cite{verma2022past}, education~\cite{singh2022metaverse}, and entertainment~\cite{dasari2023scaling}. The high spatial resolution is important to ensure the user experience because the blurry effect from the low spatial resolution would influence the engagement of viewers and might further cause discomfort. To alleviate the discomfort from the blur, current methods have proposed a series of techniques, including spatial blur~\cite{hoffman2008vergence}, defocus blur~\cite{ang2020gingervr}, depth-of-field blur~\cite{hussain2021mitigating}, and foveated rendering~\cite{meng2018kernel}. However, these methods mainly focus on the blurry regions but do not improve the local details of objects. As a result, if the viewer finds an object of interest, whatever operations s/he tries, the object always keeps blurry without any other details.

\noindent \textbf{ODI Super-Resolution.}
\cite{fakour2018360} and \cite{nishiyama2021360} take distortion maps as additional input to alleviate the distortions. LAU-Net~\cite{deng2021lau} splits an ODI into several bands along the latitude because ODIs in different latitudes present different distortions. SphereSR~\cite{yoon2022spheresr} proposes to reconstruct an ODI with arbitrary projection formats, including ERP, cube maps, and perspective formats. Recently, OSRT~\cite{yu2023osrt} and OPDN~\cite{sun2023opdn} employ transformers to construct global context and obtain good performance. However, their outputs are restricted to ERP format with no transformation.

\noindent \textbf{M\"obius Transformation.} 
It has been utilized towards straight line rectification~\cite{penaranda2018real, ferreira2017local, ferreira2020bounded}, and rotation and zoom~\cite{schleimer2016squares}. Recently,~\cite{wu2022view} employs M\"obius transformation to transform the feature maps into different forms to enhance the learning robustness. However,~\cite{wu2022view} has not explored to generate transformed ODIs with high quality. Except for ODI applications, M\"obius transformation is widely applied in data augmentation~\cite{zhou2021data}, activation function~\cite{mandic2009complex, ozdemir2011complex}, pose estimation~\cite{azizi20223d}, and convolutions~\cite{mitchel2022mobius}.
\textit{In this paper, we propose a learning-based algorithm which is built based on our conference version~\cite{cao2023omnizoomer}, to improve the textural details of ODIs when navigating and zooming in to an object of interest in VR.}

\section{OmniVR System}

\subsection{System Overview}
\label{sec: system input}
As shown in Fig.~\ref{fig:system}, we design a system, namely \textit{OmniVR} to help users effortlessly navigate and zoom in the VR meida, aiming to enhance the visual quality, and subsequently improve the immersion and interaction experience. Firstly, our system displays an original ODI for the viewer in the VR media. The viewer can freely navigate the scenario and find the object of interest. Once finding the object of interest, the viewer might feel that the object of interest is too small or not in the center of the field of view (FoV). Our system allows viewers to send commands through the VR headset and controllers. Then, the user command is utilized to generate the parameters of the M\"obius transformation matrix. Leveraging these parameters, we propose a learning-based algorithm (see Fig.~\ref{fig:4}), which is built based on our conference version~\cite{cao2023omnizoomer}, to transform the original ODI with high quality. In the end, the transformed ODI is displayed in VR to provide finer details for the viewer. Our system allows the transformed ODI with various projection formats to adapt to the visual contents in VR. Below, we describe the user command, algorithm, and view transformation of our system in detail.

\begin{figure}[t]
    \centering
    \includegraphics[width=0.8\linewidth]{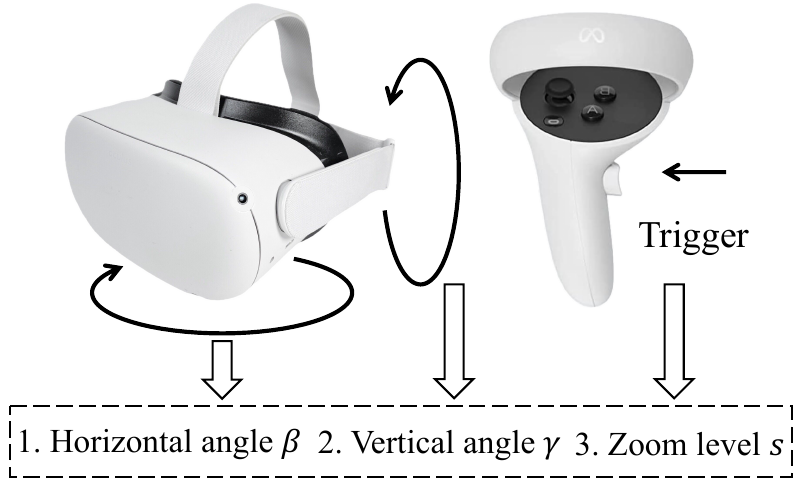}
\caption{The rotation of the VR headset generates horizontal and vertical angle parameters, while the trigger button of the right controller is used for generating zoom level parameters.}
\label{fig:user_command}
\end{figure}

\subsection{User Command and Parameter Conversion}

Our system first collects data about navigation and zoom operations through the VR headset and controllers. The rotation angle of the headset represents the navigation direction, and the trigger of the right controller controls the zoom in/out operation, as shown in Fig.~\ref{fig:user_command}. Specifically, the zoom operation is achieved by setting the UI button with arrow patterns, such as up arrows (zoom in), down arrows (zoom out), and left/right/circulation arrows (scene transition), as shown in Fig.~\ref{fig:screen_sample}. Once the raycast emitted from the controller touches the regions of UI buttons, the command is sent by clicking the trigger. Then, the collected commands are summarized with three parameters: zoom level $s$, horizontal rotation angle $\beta$, and vertical rotation angle $\gamma$, named \textit{user command}. Next, the user command is transferred to parameters $\{a,b,c,d\}$ of the M\"obius transformation matrix. We choose M\"obius transformation as it is the only bijective transformation on the sphere with preserved shape. Specifically, when performing horizontal rotation with angle $\beta$, the parameters of M\"obius transformations can be represented as follows:

\begin{equation}
\small
\begin{pmatrix}
     a & b\\
     c & d \\
\end{pmatrix} = 
\begin{pmatrix}
     \cos(\beta)+j\sin(\beta) & 0\\
     0 & 1 \\
\end{pmatrix}.
\label{eq:1}
\end{equation}

Similarly, for vertical rotation with angle $\gamma$, the parameters of M\"obius transformations can be represented as follows: 

\begin{equation}
\small
\begin{pmatrix}
     a & b\\
     c & d \\
\end{pmatrix} = 
\begin{pmatrix}
    \cos(\frac{\gamma}{2}) & \sin(\frac{\gamma}{2})\\
     -\sin(\frac{\gamma}{2}) & \cos(\frac{\gamma}{2}) \\
\end{pmatrix}.
\label{eq:2}
\end{equation}

An arbitrary rotation can be divided into horizontal rotation and vertical rotation. In addition, M\"obius transformations can be composed to give a new M\"obius transformation. Therefore, we can achieve arbitrary navigation on ODIs with horizontal rotation angle $\beta$ and vertical rotation angle $\gamma$.

For zoom with level $s$, if the pole is the North pole, the parameters of M\"obius transformations can be as follows:

\begin{equation}
\small
\begin{pmatrix}
     a & b\\
     c & d \\
\end{pmatrix} = 
\begin{pmatrix}
    s & 0 \\
     0 & 1 \\
\end{pmatrix}.
\label{eq:3}
\end{equation}

\subsection{The Proposed Algorithm}
\label{method}

\begin{figure*}[t]
    \centering
\includegraphics[width=.98\linewidth]{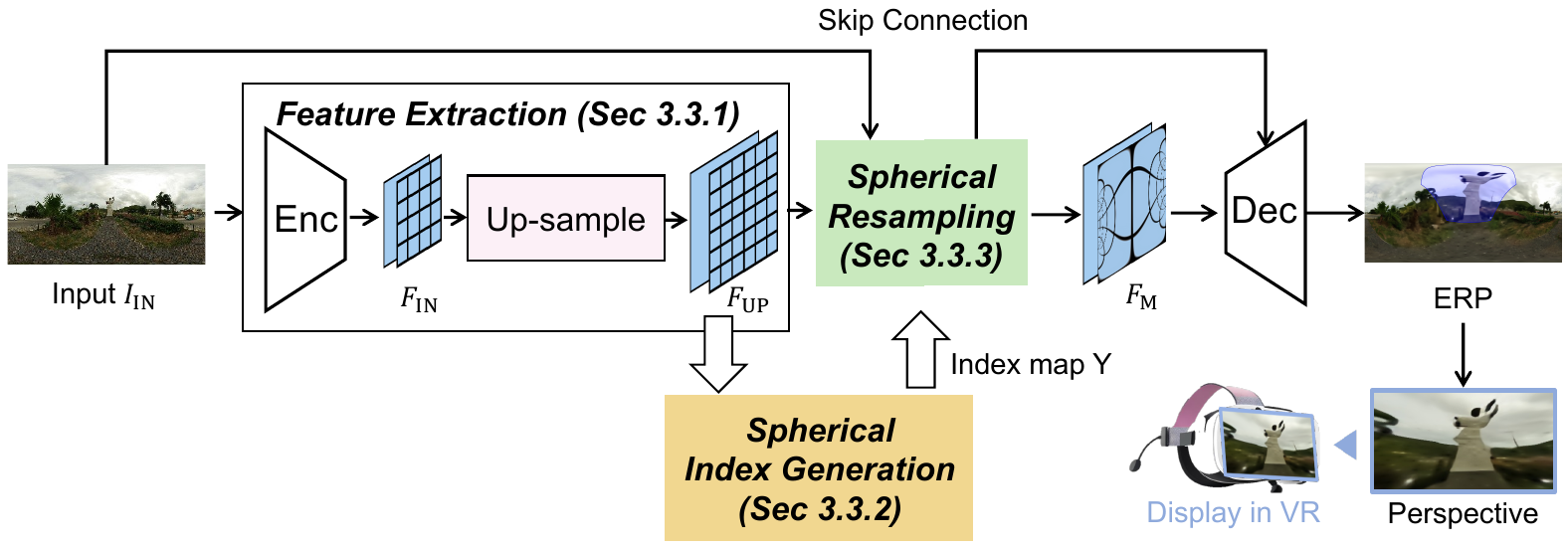}
\caption{\textbf{The overall pipeline of the proposed algorithm.} With the spatial index generation module and spherical resampling module, OmniVR can provide viewers with a flexible way to zoom in and out to objects of interest, such as the sculpture.}
\label{fig:4}
\end{figure*}

\noindent{\textbf{Overview.}} As depicted in Fig.~\ref{fig:4}, we propose a novel algorithm that allows for free navigation to objects of interest and zooming in with preserved shapes and high-quality details, based on our OmniZoomer~\cite{cao2023omnizoomer}. Initially, we extract HR feature maps $F_{\text{UP}} \in \mathbb{R}^{H\times W \times C}$ from the input ODI $I_{\text{IN}} \in \mathbb{R}^{h\times w \times 3}$ through an encoder and an up-sampling block (Sec.~\ref{sec:upsample}). With $F_{\text{UP}}$'s index map $X \in \mathbb{R}^{H\times W \times 2}$ as the input, we propose the spatial index generation module (Sec.~\ref{sec:grid}) to apply the M\"obius transformation~\cite{Kato2015MobiusTA} according to the user command on $X$ to generate the transformed spatial index map $Y \in \mathbb{R}^{H\times W \times 2}$.
Note that the channel numbers of $X$ and $Y$ indicate the longitude and latitude, respectively. Subsequently, we introduce a spherical resampling module (Sec.~\ref{sec:Slerp}) that generates the transformed HR feature maps $F_{\text{M}} \in \mathbb{R}^{H\times W \times C}$ by resampling the pixels on the sphere guided by $Y$. Finally, we decode the transformed feature maps to output the transformed ODI. The decoder consists of three ResBlocks~\cite{lim2017enhanced} and a convolution layer. We take the same parameters in the spatial index generation module to transform the HR ground truth ODIs and employ the $L1$ loss for supervision. The proposed algorithm in OmnIVR has two main differences with OmniZoomer~\cite{cao2023omnizoomer}: 1) To stabilize the convergence during training, we employ the skip connection by applying M\"obius transformation onto $I_{\text{IN}}$, which is added to the output of the decoder. 2) The proposed algorithm in our OmniVR enjoys arbitrary spherical projection, \ie, ERP and perspective projection, to meet the demands of viewing ODIs in VR with different FoVs. We now provide detailed descriptions of these components. 

\subsubsection{Feature Extraction}
\label{sec:upsample}

Given an ODI $I_{\text{IN}}\in\mathbb{R}^{h\times w \times 3}$ in ERP format, our initial step involves the use of an encoder composed of several convolution layers. This encoder is to extract the feature maps $F_{\text{IN}}\in\mathbb{R}^{h\times w \times C}$. Subsequently, we employ an upsampling block, equipped with multiple pixel-shuffle layers [\cite{Shi2016RealTimeSI}], to produce HR feature maps $F_{\text{UP}}\in\mathbb{R}^{H\times W \times C}$. Here, $H=s*h$, $W=s*w$ represent the spatial dimensions of the up-sampled image, where $s$ denotes the scale factor and $C$ indicates the number of channels. Notably, we apply the M\"obius transformation in the HR space to address the aliasing issue. This issue arises due to inadequate pixel representation for accurately describing continuous curves post-transformation, potentially leading to object shape distortion.

\subsubsection{Spatial Index Generation }
\label{sec:grid}

We apply the M\"obius transformation on the spatial index map $X$ of HR feature maps $F_{\text{UP}}$ and generate the transformed spatial index map $Y$ for the subsequent resampling operation. M\"obius transformation is known as the only conformal bijective transformation between the sphere and the complex plane. To apply the M\"obius transformation on the HR feature maps $F_{\text{UP}}$, we first use spherical projection (SP) to project the spatial index map $X$ from spherical coordinates $(\theta, \phi)$ (where $\theta$ represents the longitude and $\phi$ represents the latitude) to the Riemann sphere ${\mathbb {S}^2 = \{(x,y,z) \in \mathbb {C}^3 | x^2+y^2+z^2=1}\}$, formulated as:
\begin{equation}
\text{SP}: \begin{pmatrix}
     x  \\
     y \\
     z \\
\end{pmatrix} = 
\begin{pmatrix}
     \cos(\phi)\cos(\theta) \\
     \cos(\phi)\sin(\theta) \\
     \sin(\phi) \\
\end{pmatrix} .
\label{eq:4}
\end{equation}
Then, with stereographic projection (STP)~\cite{Eybpoosh2021ApplyingIS}, we can project a point $(x, y, z)$ of the Riemann sphere $\mathbb {S}^2$ onto the complex plane and obtain the projected point ($x'$, $y'$). Let point $(0,0,1)$ be the pole, STP can be formulated as:
{\begin{equation}
\text{STP}: x' = {\frac {x }{1 - z}}\ ,\ y' = {\frac {y}{1 - z}}.
\label{eq:5}
\end{equation}}

Subsequently, given the projected point $p$ ($Z_p$ = $x'$+$i y'$) on the complex plane, we can conduct the M\"obius transformation with the following formulation:
{\begin{equation}
f(Z_p)={\frac {aZ_p+b}{cZ_p+d}}, 
\label{eq:6}
\end{equation}}
where $a$, $b$, $c$, and $d$ are complex numbers satisfying $ad-bc \neq 0$. Finally, we apply the inverse stereographic projection $\text{STP}^{-1}$ and inverse spherical projection $\text{SP}^{-1}$ to re-project the complex plane into the ERP plane:
\begin{equation}
\begin{split}
\renewcommand{\arraystretch}{1.5}
\setlength{\arraycolsep}{1.3pt}
    \text{STP}^{-1}: \begin{pmatrix}
         x  \\
         y \\
         z \\
    \end{pmatrix} &= 
    \begin{pmatrix}
         \frac{2x'}{1+x'^2+y'^2} \\
         \frac{2y'}{1+x'^2+y'^2} \\
         \frac{-1+x'^2+y'^2}{1+x'^2+y'^2} \\
        \end{pmatrix} \ ; \\
   \ \text{SP}^{-1}: \begin{pmatrix}
         \theta \\
         \phi \\
    \end{pmatrix} &= 
    \begin{pmatrix}
         \arctan(y/x) \\
         \arcsin(z) \\
        \end{pmatrix} \ .
\end{split}
    \label{eq:7}
\end{equation}

\begin{figure}[t]
    \centering
\includegraphics[width=.85\linewidth]{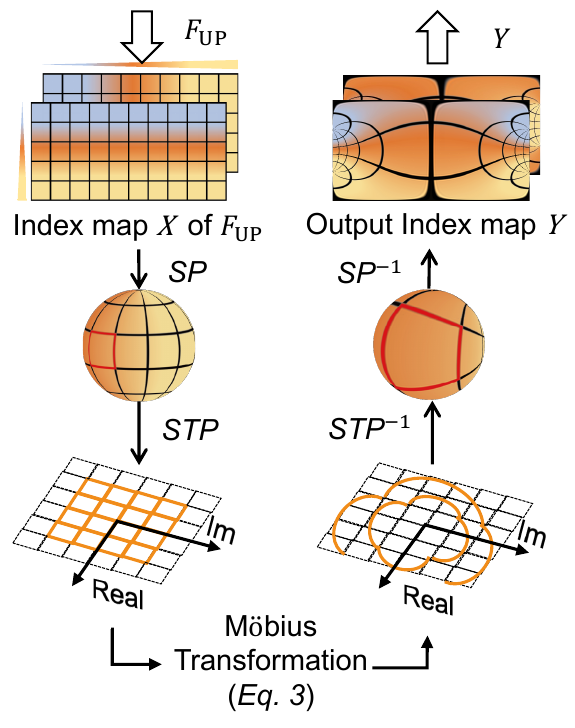}
\caption{\textbf{The illustration of the proposed spatial index generation module.} With the HR feature map as input, the spatial index generation module generates a transformed index map to accomplish the feature transformation process.}
\label{fig:mobius}
\end{figure}

In summary, as shown in Fig.~\ref{fig:mobius}, we first project the index map $X$ of input feature $F_{\text{UP}}$ to the complex plane using SP (Eq.~\ref{eq:4}) and STP (Eq.~\ref{eq:5}), and then conduct the M\"obius transformation with Eq.~\ref{eq:6}, and generate the transformed index map $Y$ through the inverse STP (Eq.~\ref{eq:7}) and inverse SP (Eq.~\ref{eq:7}).

\subsubsection{Spherical Resampling}
\label{sec:Slerp}

\begin{figure}[t]
    \centering
    \includegraphics[width=0.9\linewidth]{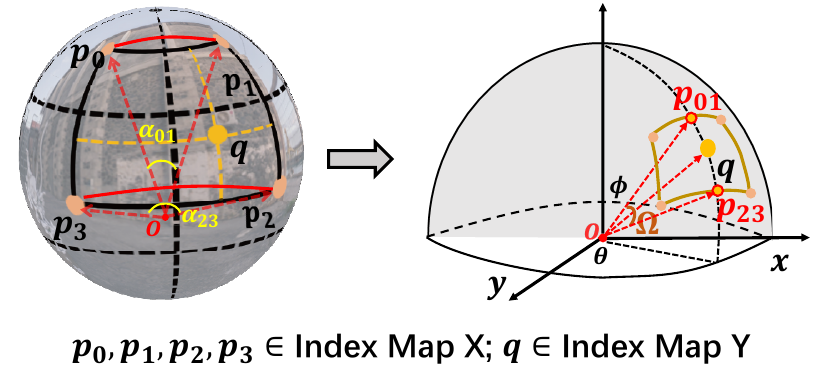}
\caption{Spherical resampling considers the angles (\ie, $\alpha_{01}$, $\alpha_{23}$, $\Omega$) between points on the sphere, which correspond to the red solid curves.}
\vspace{-15pt}
\label{fig:7}
\end{figure}

Inspired by the inherent spherical representation of ODIs and the spherical conformality of M\"obius transformation, we propose the spherical resampling module to generate the transformed feature maps $F_\text{M}$. The spherical resampling module resamples on the curved sphere based on the spherical geodesic of two points on the sphere.
Given a query pixel $q$ with the spatial index $(\theta_q, \phi_q)$ from the index map $Y$, we choose its four corner pixels $\{p_i, i=0,1,2,3\}$ as the neighboring pixels, which are located on the feature maps $F_\text{UP}$ (as shown in the left of Fig.~\ref{fig:7}. 
The indices of the neighboring pixels satisfy the following conditions: $\theta_0=\theta_3$, $\theta_1=\theta_2$, $\phi_0=\phi_1$, and $\phi_2=\phi_3$. To obtain the feature value of the query pixel $q$, we employ the spherical linear interpolation (Slerp)~\cite{Fatelo2021MobilitySA}, 
which is a constant-speed motion along the spherical geodesic of two points on the sphere, formulated as follows:
{\begin{equation}
    \text{Slerp}(a,b) = \frac{\sin(1-t) \beta}{\sin\beta}a + \frac{\sin t \beta}{\sin\beta}b,
    \label{eq:slerp}
\end{equation}}
where $\beta$ is the angle subtended by $a$ and $b$, and $t$ is the resampling weight. Note that $t$ is easy to determine if $a$ and $b$ are located on the same longitude. Therefore, we calculate the feature value of pixel $q$ with two steps. Firstly, we resample $p_0,p_1$ and $p_2,p_3$ to $p_{01}$ and $p_{23}$, respectively, as shown in the right of Fig.\ref{fig:7}. Taking the resampling of $p_{0,1}$ as example, the formulation can be described as:
{\begin{equation}
    F(p_{01}) = \frac{\sin(1-t_{01}) \alpha_{01}}{\sin \alpha_{01}}F(p_{0}) + \frac{\sin t_{01} \alpha_{01}}{\sin \alpha_{01}}F(p_{1}),
    \label{eq:9}
\end{equation}}

\begin{figure*}[t]
    \centering
    \includegraphics[width=0.94\textwidth]{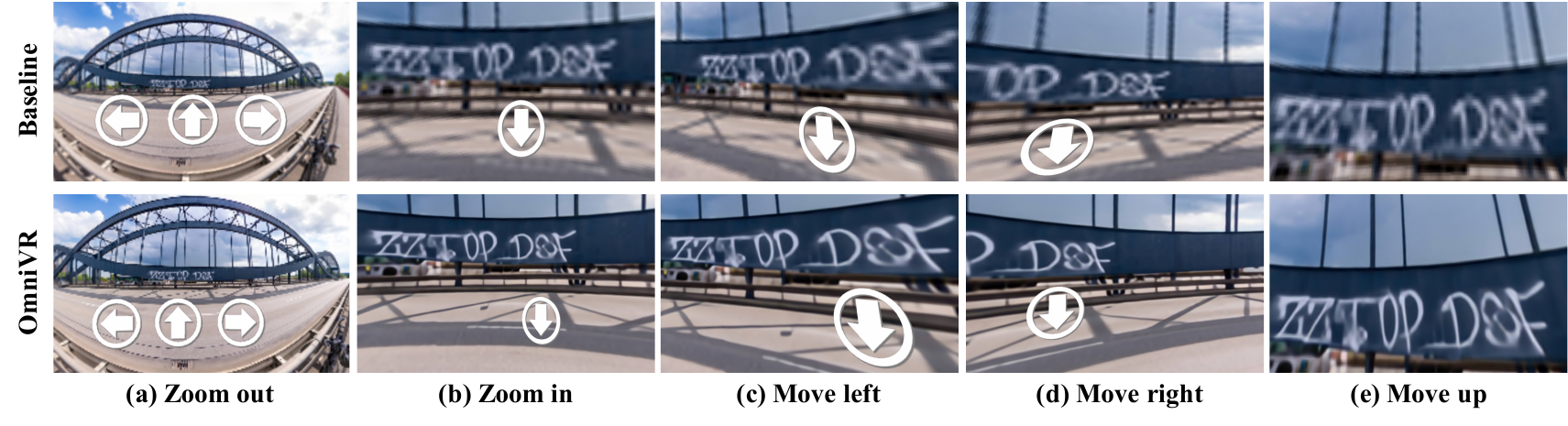}
\caption{\textbf{The samples in the third scenario.} The participant can freely choose the viewing direction and zoom level to get a better viewing experience.}
\label{fig:screen_sample}
\end{figure*}

\noindent where $\alpha_{01}$ is the angle subtended by $p_0$ and $p_1$, and the weight $t_{01}$ is decided by the location of $p_{01}$ on the curve $\overset{\LARGE{\frown}}{p_{0}p_{1}}$. Notably, $t_{01}$ should ensure $p_{01}$ to have the same longitude with the query pixel $q$. Similarly, $\alpha_{23}$ is the angle subtended by $p_2$ and $p_3$, and $p_{23}$ also has the same longitude with the query pixel $q$ by calculating the weight $t_{23}$. After that, we follow the Slerp (Eq.~\ref{eq:slerp}) to calculate the feature value $F_{q}$ as follows:
{\begin{equation}
F(q)=\frac{\sin(1-t_q)\Omega}{\sin\Omega}F(p_{01}) + \frac{\sin t_q\Omega}{\sin\Omega}F(p_{23}),
\label{eq:10}
\end{equation}}

\noindent where $\Omega$ is the angle subtended by $p_{01}$ and $p_{23}$, and $t_q$ is decided by the location of $q$ on the curve $\overset{\LARGE{\frown}}{p_{01}p_{23}}$. If we assume that $p_0$, $p_1$, and $p_{01}$ have the same latitude, the calculation of $t_{01}$ can be simplified into $\frac{\theta_q - \theta_0}{\theta_1 - \theta_0}$. Similarly, $t_{23}$ can be simplified into $\frac{\theta_q - \theta_2}{\theta_3 - \theta_2}$. Also, $t_{q}$ can be simplified into $\frac{\phi_q - \phi_{01}}{\phi_{23} - \phi_{01}}$.

\subsection{View Transformation in VR}
\label{sec: system output}

As shown in Fig.~\ref{fig:screen_sample},  the transformed ODI is displayed by our proposed system. As OmniZoomer~\cite{cao2023omnizoomer} only supports generating ODIs with ERP format, its output can not meet the demands of viewing ODIs in VR with different FoVs. In contrast, the proposed algorithm in our OmniVR allows for various projection formats by adding a projection transformation layer. In this case, the transformed ODI can be displayed in VR with the perspective format that fits the FoV of the user with less distortion effect. For detailed projection transformations, we recommend readers to the recent survey~\cite{ai2022deep}. The transformed ODI by our algorithm contains finer details, which could assist the user in recognizing and understanding the scenario, thus improving the immersive and interactive experience in VR. As shown in Fig.~\ref{fig:screen_sample}, our system allows displaying the generated HR ODIs with perspective views in the VR media under various user commands, including zooming in/out and moving towards different directions, \eg, left, right, and up.

\section{Experiment for Algorithm}
\label{experiment}

\subsection{Dataset and Implementation Details}
\label{dataset}

\noindent \textbf{Datasets.} No datasets exist for ODIs under M\"obius transformations, and collecting real-world ODI pairs with corresponding M\"obius transformation matrices is difficult.  Thus, we propose ODI-M\"obius (ODIM) dataset to train our OmniVR and compare methods in a supervised manner. Our dataset is based on the ODI-SR dataset~\cite{deng2022omnidirectional} with 1191 images in the train set, 100 images in the validation set, and 100 images in the test set. Note that the proposed dataset is consistent with the conference version~\cite{cao2023omnizoomer}. During training, as our system aims to freely navigate and zoom in the VR media, we generate user commands that include horizontal rotation, vertical rotation, and zoom-in/out operations. The user commands are converted to the parameters $\{a,b,c,d\}$ of the M\"obius transformation matrix according to Eq.~\ref{eq:6}. During validating and testing, we assign a fixed user command and the corresponding M\"obius transformation matrix for each ODI. Besides, we test on SUN360~\cite{Xiao2012RecognizingSV} dataset with 100 images.

\begin{table*}[t!]
\centering
\setlength{\tabcolsep}{2.6mm}{
\begin{tabular}{c||cccc||cccc}
\toprule
Scale & \multicolumn{4}{c||}{$\times 8$} & \multicolumn{4}{c}{$\times 16$} \\ \hline
\multirow{2}{*}{Method} & \multicolumn{2}{c|}{ODI-SR} & \multicolumn{2}{c||}{SUN 360} & \multicolumn{2}{c|}{ODI-SR} & \multicolumn{2}{c}{SUN 360} \\ \cline{2-9} & \multicolumn{1}{c|}{WS-PSNR} & \multicolumn{1}{c|}{WS-SSIM} & \multicolumn{1}{c|}{WS-PSNR} & WS-SSIM & \multicolumn{1}{c|}{WS-PSNR} & \multicolumn{1}{c|}{WS-SSIM} & \multicolumn{1}{c|}{WS-PSNR} & WS-SSIM \\ \hline \hline
Bicubic  & \multicolumn{1}{c|}{26.77} & \multicolumn{1}{c|}{0.7725} & \multicolumn{1}{c|}{25.87} & \multicolumn{1}{c||}{0.7103} & \multicolumn{1}{c|}{24.79} & \multicolumn{1}{c|}{0.7404} & \multicolumn{1}{c|}{23.87} & \multicolumn{1}{c}{0.6802} \\ \hline
RCAN${\rm (+Transform)}$~\cite{zhang2018image}  & \multicolumn{1}{c|}{27.46} & \multicolumn{1}{c|}{0.7906} & \multicolumn{1}{c|}{27.04} & \multicolumn{1}{c||}{0.7443} & \multicolumn{1}{c|}{25.45} & \multicolumn{1}{c|}{0.7541} & \multicolumn{1}{c|}{24.70}  & \multicolumn{1}{c}{0.7001}  \\ \hline
LAU-Net${\rm (+Transform)}$~\cite{deng2021lau} & \multicolumn{1}{c|}{27.25} & \multicolumn{1}{c|}{0.7813}       & \multicolumn{1}{c|}{26.77}     & \multicolumn{1}{c||}{0.7363} & \multicolumn{1}{c|}{25.23} & \multicolumn{1}{c|}{0.7455} & \multicolumn{1}{c|}{24.49} & \multicolumn{1}{c}{0.6921} \\ \hline
OmniZoomer-RCAN  & \multicolumn{1}{c|}{27.53} & \multicolumn{1}{c|}{0.7970}& \multicolumn{1}{c|}{27.34} & \multicolumn{1}{c||}{0.7592} & \multicolumn{1}{c|}{25.50} & \multicolumn{1}{c|}{0.7584}  & \multicolumn{1}{c|}{24.84} & \multicolumn{1}{c}{0.7034} \\ \hline
OmniVR-RCAN  & \multicolumn{1}{c|}{\textbf{27.62}} & \multicolumn{1}{c|}{\textbf{0.8005}}& \multicolumn{1}{c|}{\textbf{27.50}} & \multicolumn{1}{c||}{\textbf{0.7662}} & \multicolumn{1}{c|}{\textbf{25.52}} & \multicolumn{1}{c|}{\textbf{0.7629}}  & \multicolumn{1}{c|}{\textbf{24.89}} & \multicolumn{1}{c}{\textbf{0.7094}} \\ \bottomrule
\end{tabular}}
\caption{\textbf{Quantitative comparison of M\"obius transformation results on ODIs.} ${\rm (+Transform)}$ denotes that we first employ a scale-specific SR model for image SR and then conduct image-level M\"obius transformation on the SR image. We report on ODI-SR dataset and SUN360 dataset with up-sampling factors $\times 8$ and $\times 16$. \textbf{Bold} indicates the best results.}
\label{tab:msr}
\end{table*}

\begin{table*}[t!]
\centering
\setlength{\tabcolsep}{3.5mm}{
\begin{tabular}{c||cccc||cccc}
\toprule
Scale & \multicolumn{4}{c||}{$\times 8$} & \multicolumn{4}{c}{$\times 16$} \\ \hline
\multirow{2}{*}{Method} & \multicolumn{2}{c|}{ODI-SR} & \multicolumn{2}{c||}{SUN 360} & \multicolumn{2}{c|}{ODI-SR} & \multicolumn{2}{c}{SUN 360} \\ \cline{2-9} & \multicolumn{1}{c|}{WS-PSNR} & \multicolumn{1}{c|}{WS-SSIM} & \multicolumn{1}{c|}{WS-PSNR} & WS-SSIM & \multicolumn{1}{c|}{WS-PSNR} & \multicolumn{1}{c|}{WS-SSIM} & \multicolumn{1}{c|}{WS-PSNR} & WS-SSIM \\ \hline \hline
Bicubic  & \multicolumn{1}{c|}{19.64} & \multicolumn{1}{c|}{0.5908} & \multicolumn{1}{c|}{19.72} & \multicolumn{1}{c||}{0.5403} & \multicolumn{1}{c|}{17.12} & \multicolumn{1}{c|}{0.4332} & \multicolumn{1}{c|}{17.56} & \multicolumn{1}{c}{0.4638} \\ \hline
EDSR~\cite{lim2017enhanced} & \multicolumn{1}{c|}{23.97} & \multicolumn{1}{c|}{0.6483} & \multicolumn{1}{c|}{23.79} & \multicolumn{1}{c||}{0.6472} & \multicolumn{1}{c|}{22.24} & \multicolumn{1}{c|}{0.6090} & \multicolumn{1}{c|}{21.83} & \multicolumn{1}{c}{0.5974} \\ \hline 
RCAN~\cite{zhang2018image} & \multicolumn{1}{c|}{24.26} & \multicolumn{1}{c|}{0.6554} & \multicolumn{1}{c|}{23.88} & \multicolumn{1}{c||}{0.6542}        & \multicolumn{1}{c|}{22.49}   & \multicolumn{1}{c|}{0.6176} & \multicolumn{1}{c|}{21.86}        & \multicolumn{1}{c}{0.5938} \\ \hline
360-SS~\cite{ozcinar2019super} & \multicolumn{1}{c|}{24.14} & \multicolumn{1}{c|}{0.6539} & \multicolumn{1}{c|}{24.19} & \multicolumn{1}{c||}{0.6536} & \multicolumn{1}{c|}{22.35} & \multicolumn{1}{c|}{0.6102} & \multicolumn{1}{c|}{22.10} & \multicolumn{1}{c}{0.5947} \\ \hline
SphereSR~\cite{yoon2022spheresr} & \multicolumn{1}{c|}{24.37} & \multicolumn{1}{c|}{0.6777} & \multicolumn{1}{c|}{24.17} & \multicolumn{1}{c||}{0.6820} & \multicolumn{1}{c|}{22.51} & \multicolumn{1}{c|}{0.6370} & \multicolumn{1}{c|}{21.95} & \multicolumn{1}{c}{0.6342} \\ \hline
LAU-Net~\cite{deng2021lau} & \multicolumn{1}{c|}{24.36} & \multicolumn{1}{c|}{0.6602} & \multicolumn{1}{c|}{24.24} & \multicolumn{1}{c||}{0.6708}  & \multicolumn{1}{c|}{22.52} & \multicolumn{1}{c|}{0.6284} & \multicolumn{1}{c|}{22.05} & \multicolumn{1}{c}{0.6058} \\ \hline 
LAU-Net+~\cite{deng2022omnidirectional} & \multicolumn{1}{c|}{\textbf{24.63}} & \multicolumn{1}{c|}{0.6815} & \multicolumn{1}{c|}{24.37} & \multicolumn{1}{c||}{0.6710} & \multicolumn{1}{c|}{\textbf{22.97}} & \multicolumn{1}{c|}{0.6316} & \multicolumn{1}{c|}{\textbf{22.22}} & \multicolumn{1}{c}{0.6111} \\ \hline
OmniVR-RCAN  & \multicolumn{1}{c|}{24.61} & \multicolumn{1}{c|}{\textbf{0.6822}} & \multicolumn{1}{c|}{\textbf{24.53}} & \multicolumn{1}{c||}{\textbf{0.7152}} & \multicolumn{1}{c|}{22.68}  & \multicolumn{1}{c|}{\textbf{0.6324}} & \multicolumn{1}{c|}{22.14} & \multicolumn{1}{c}{\textbf{0.6483}} \\ \bottomrule
\end{tabular}}
\caption{\textbf{Quantitative comparison of ODI SR task.} The numbers are excerpted from ~\cite{deng2022omnidirectional} except for~\cite{yoon2022spheresr}, due to its reported results are obtained by utilizing 800 training images in the ODI-SR dataset. We report $\times 8$, $\times 16$ SR results on the ODI-SR and SUN360 datasets. Bold indicates the best.}
\label{tab:sr}
\end{table*}



\noindent \textbf{Implementation details.} We mainly evaluate the ERP format. The resolution of the HR ERP images is $1024 \times 2048$, and the up-sampling factors we choose are $\times8$ and $\times16$. We use L1 loss, which is optimized by
Adam optimizer~\cite{kingma2014adam}, with an initial learning rate of 1e-4. The batch size is 1 when using RCAN~\cite{zhang2018image} as the backbone. Especially, considering the spherical imagery of ODIs, we use specific WS-PSNR~\cite{sun2017weighted} and WS-SSIM~\cite{zhou2018weighted} metrics for evaluation.

\begin{figure*}[t]
    \centering
    \includegraphics[width=0.9\linewidth]{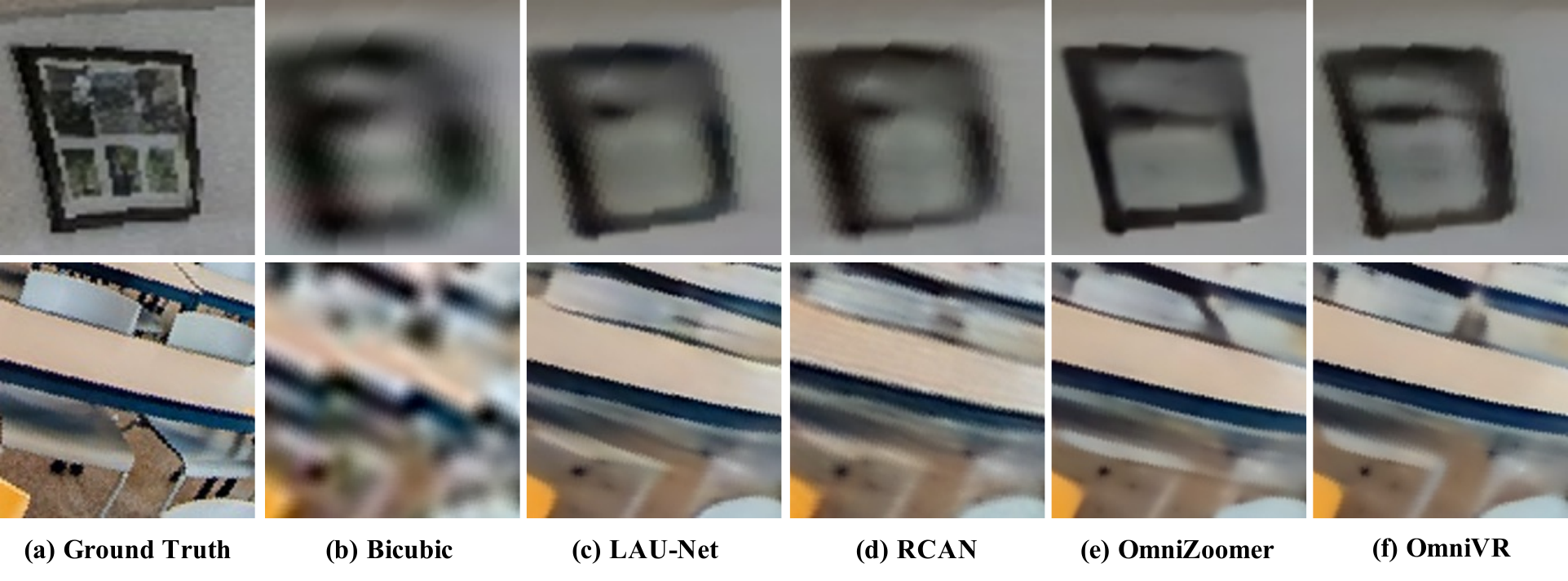}
\caption{Visual comparisons of M\"obius transformation results on ODI-SR (1st row) and SUN360 (2nd row) datasets.}
\label{fig:comparex8}
\end{figure*}

\subsection{Quantitative and Qualitative Evaluation}
\label{comparison}

\noindent \textbf{Navigate and Zoom in:}
Except for OmniZoomer~\cite{cao2023omnizoomer}, there are no prior arts that can be directly compared. For a fair comparison,  we combine the existing image SR models~\cite{zhang2018image,deng2021lau} with image-level M\"obius transformations. The SR models designed for 2D planar images are retrained based on their official settings.

In Tab.~\ref{tab:msr}, by applying RCAN as the backbone, OmniVR outperforms current methods in all metrics, all up-sampling factors, and test sets. 
It reveals the effectiveness of our OmniVR incorporating M\"obius transformation into the neural network. Note that LAU-Net shows inferior performance because it is limited to vertically-captured ODIs. Importantly, our OmniVR outperforms OmniZoomer in all metrics, demonstrating the importance of skip connection for stable convergence during training. As shown in the first row of Fig.~\ref{fig:comparex8} on the ODI-SR dataset, OmniVR predicts clearer picture frames. Similarly, in the second row of Fig.~\ref{fig:comparex8} on the SUN360 dataset, OmniVR reconstructs clearer structures of the chairs than other methods.

\noindent \textbf{Direct SR:} Our OmniVR can achieve the naive SR task by setting the M\"obius transformation matrix as an identity matrix. Tab.~\ref{tab:sr} shows that OmniVR with RCAN as backbone obtains 5 (total 8) best metrics. It demonstrates the strong capability of our method to handle the inherent distortions.

\begin{figure*}[t]
    \centering
    \includegraphics[width=0.8\linewidth]{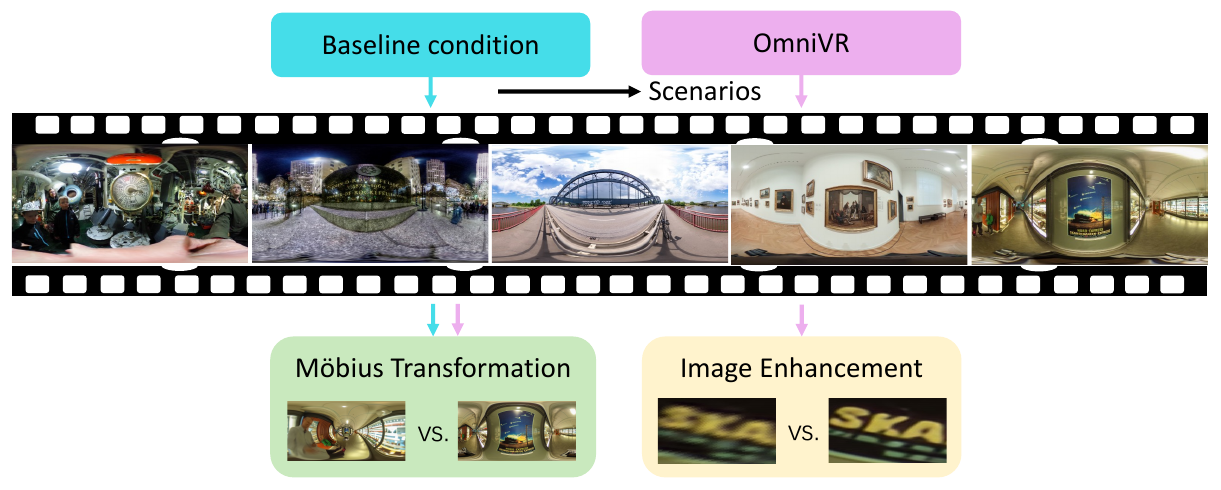}
\caption{Overview of the five scenarios and their order. The user study contains the baseline condition and OmniVR. The two techniques both use M\"obius transformation for zoom in/out, while OmniVR additionally combines image enhancement.}
\label{fig:experiment_setting}
\end{figure*}

\begin{figure}[t]
    \centering
    \includegraphics[width=0.9\linewidth]{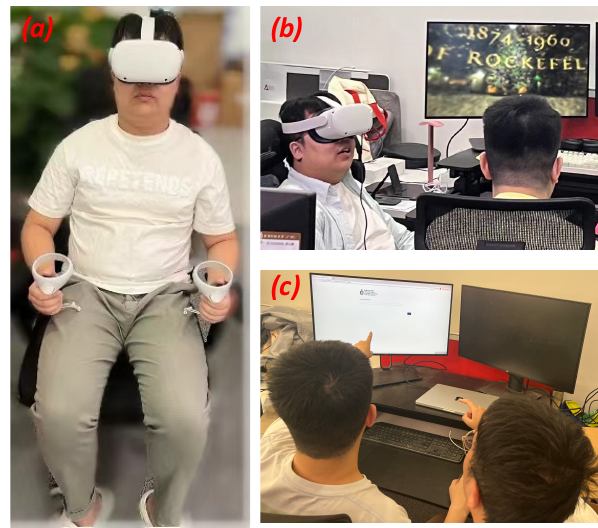}
\caption{Sample scenes during user study. (a) The participant is equipped with the VR headset and controllers. (b) The participant is viewing the scenarios, which are displayed on the screen simultaneously. (c) Before and after equipping the VR headsets, the participant fills the basic information and interviews, respectively.}
\label{fig:user}
\end{figure}


\section{User Study}

We conduct a within-subject user study in VR to explore the effectiveness of OmniVR and the overall user experience of our proposed system. Note that the user study involves no more than minimal risk, and the IRB board has granted a waiver for the review process.
To establish a comparative baseline for assessing the image enhancement performance of OmniVR, we include a condition that only enables M\"obius transformation with Bicubic interpolation in Fig.~\ref{fig:experiment_setting}.

\subsection{Experiment Set-Up \& Participants}

\noindent \textbf{Participants.} We recruited 18 participants (P1-P18) through the university mailing list, including 8 males and 10 females. 56\% of them are between 18 and 24 years old, and 44\% of them are between 25 and 34 years old. 9 participants have viewed ODIs on VR devices and mobile phones, and 6 participants have viewed ODIs on computers. Furthermore, their familiarity with 3D games or 3D models is mandatory (if 1 represents very low and 7 represents very high, the average is 3.94). Each participant received 3 dollars as compensation.

\noindent \textbf{Appratus \& Data.}
We employ a Meta Quest 2 for the experiment, as shown in Fig.~\ref{fig:user}(a). We select five ODIs as five scenarios in the VR experience. The ODIs are from the training and testing sets of the Flickr360 dataset~\cite{cao2023ntire}. The selected ODIs have texts or textures in the equator regions, which have a distinction in different zoom levels and spatial resolutions. The ODIs are various from indoor scenarios to outdoor scenarios. The transition among different scenarios and zooming in/out are achieved using the CenarioVR software~\footnote{https://www.cenariovr.com/} and controlled through specific arrow keys. 
During the VR experience, the view direction is controlled through the head movements, while the scenario transition is controlled with the trigger of the right VR controller.

\begin{figure}[t]
    \centering
    \includegraphics[width=\linewidth]{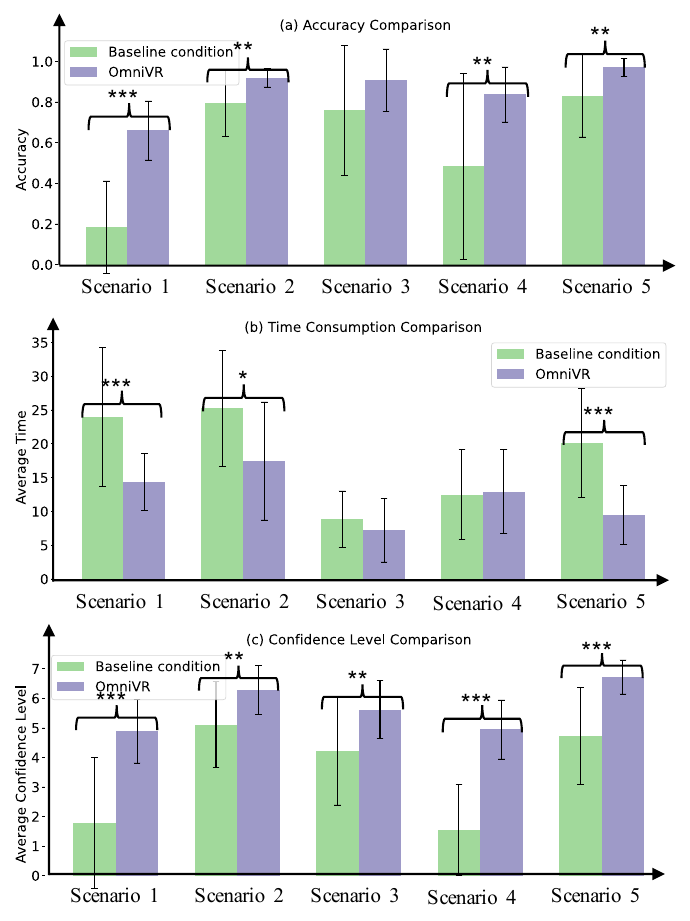}
\caption{Means and standard deviations of five scenarios from two groups, \ie, the baseline condition and OmniVR. We also present the difference between the two groups using a t-test.}
\label{fig:mean_std}
\end{figure}

\noindent \textbf{Experimental Conditions.} As shown in Fig.~\ref{fig:experiment_setting}, we set a baseline condition for comparison. The baseline condition enables M\"obius transformation that enables zoom-in/out for users to find details in various zoom levels. Differently, the M\"obius transformation in the baseline condition is conducted with Bicubic interpolation, while in OmniVR it is achieved by deep learning. Moreover, OmniVR additionally contains image enhancement to recover more details.

\subsection{Design \& Procedure}
The experiment was within-subjects: 2 technique $\times$ 5 scenarios $\times$ 3 answer ranges (\ie, accuracy, confidence, and response time) = 30 responses per participant.
We fixed the order of two conditions because ODIs obtained from OmniVR exhibit higher quality than the baseline condition in Sec.~\ref{experiment}. Firstly viewing ODIs generated with OmniVR would cause information leakage to the following baseline condition and thus influence the task performance from OmniVR. To further avoid the influence of this memorization issue when comparing two conditions, we design two questions with similar difficulty in each scenario. These questions are about text recognition in four scenarios, and number counting of a specific object in one scenario. Participants would respond to one question in the baseline condition, and respond to the other question in the OmniVR condition. The order of the two questions is random, while the number of participants receiving some order is consistent.
We also fixed the order of scenarios because there is no relationship between them.
The experiment lasted for about 20 minutes on average. 

\begin{figure}[t]
    \centering
    \includegraphics[width=0.9\linewidth]{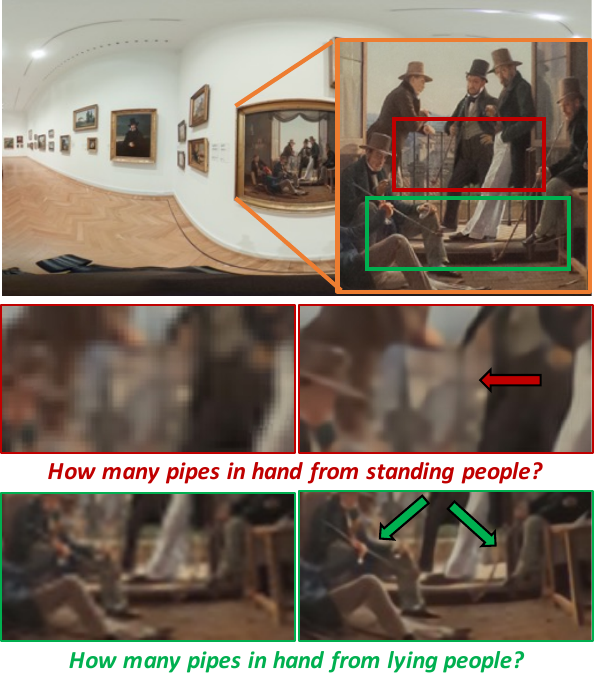}
\caption{Visualization of scenario 4. The first row is the HR ground truth ODI. The second and third rows represent the areas related to the two questions, respectively. The left patches are generated with the baseline condition, while the right patches are generated with OmniVR.}
\label{fig:none_visual}
\end{figure}

This study contains two parts.
Before starting the study, each participant was given a short introduction to the study and our system. 
The VR headset and controllers were adjusted for each participant to ensure that the testing text was clear.
The first part focused on assessing the image enhancement performance of our proposed technique.
The participants first viewed the five scenarios generated with the baseline condition and then viewed them generated with OmniVR, as shown in Fig.~\ref{fig:user}(b). 
This order would help participants forget specific impressions for a fair comparison.
For each technique $\times$ scenario, participants were asked to answer the question as quickly as possible.
The second part focused on the overall user experience of the whole system.
We asked participants to fill in a 7-point Likert questionnaire to measure their cybersickness, mental and physical costs, immersive experience, and usability of zoom-in techniques for two techniques, as shown in Fig.~\ref{fig:user}(c).
For a fair comparison, the participants were only informed that the first and five scenarios are generated with "Technique 1" and "Technique 2" respectively.
We also conduct a post-study interview to collect user feedback about the suggestions and expectations for our technique and system.

\subsection{Measures}
In the first part, we set three metrics for evaluation: accuracy, response time, and confidence level. Ideally, higher accuracy, shorter response time, and higher confidence levels could demonstrate higher image quality. We set questions about text recognition and number counting for a specific object. For the text recognition question, the accuracy is the ratio of correct words. For the number counting question, the accuracy is calculated through $e^{-|N_1 - N_2|}$, where $N_1$ and $N_2$ are the responded and right numbers, respectively. The confidence level is measured within a 7-point Likert scale. In addition, we allow participants to respond ``Invisible" if the scenario is too blurry to identify. In this case, the accuracy and confidence level are set to zero.  In the second part, we collected their 7-point Likert scale ratings about cybersickness, mental and physical costs, immersive experience, and usability (1 -- very low, 7 -- very high). 

\subsection{Results \& Findings}
\subsubsection{Quantitative Evaluation}

We utilize a t-test to analyze the participants' VR experience in five scenarios. The main reason is that we want to compare the difference between the baseline condition and OmniVR. 

\begin{figure}[t]
    \centering
    \includegraphics[width=\linewidth]{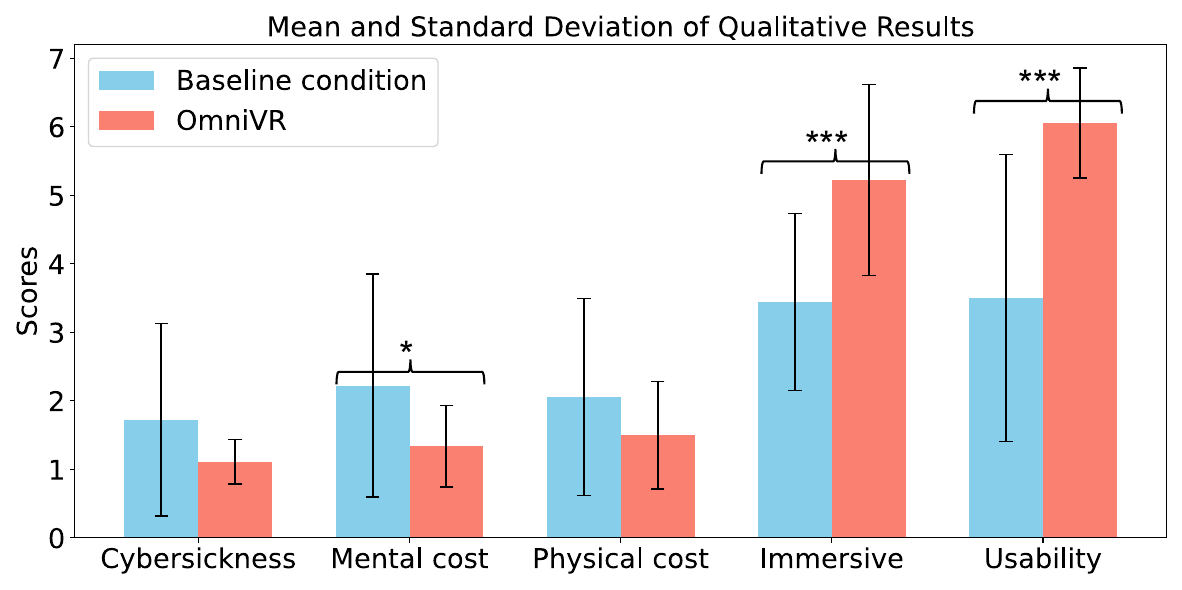}
\caption{Qualitative results of user study, including cybersickness, mental and physical costs, immersion, and usability. We also present the difference between the two groups using a t-test.}
\label{fig:interview}
\end{figure}

In Fig.~\ref{fig:mean_std}, we report the comparison results between the baseline condition and OmniVR across five scenarios and three metrics. For accuracy in Fig.~\ref{fig:mean_std}(a), OmniVR surpasses the baseline condition consistently in five scenarios. Especially, in scenario 1, OmniVR outperforms 0.48 compared with the baseline condition. This is mainly because in scenario 1 the baseline condition recovers few local details, and thus 5 participants respond ``Invisible". Similar occasion occurs in scenario 4, where 8 participants respond ``Invisible" under the baseline condition. For response time in Fig.~\ref{fig:mean_std}(b), OmniVR consumes less time than the baseline condition in most of the scenarios and comparable time with the baseline condition in scenario 4. For confidence level in Fig.~\ref{fig:mean_std}(c), OmniVR obtains higher confidence levels in all scenarios. It demonstrates that OmniVR helps participants identify the local details easily through zoom-in functions and image enhancement. In this case, the ambiguous occasions that degrade the confidence level are reduced significantly.

We further utilize a t-test to explore the difference between the baseline condition and OmniVR. In Fig.~\ref{fig:mean_std}, the baseline condition and OmniVR show significant differences in most scenarios and metrics. For scenario 3, the accuracy and response time between the two conditions have no significant difference. We ascribe it as the simple and easily identifiable words in scenario 3. For scenario 4, the response time between the two groups shows no significant difference. In Fig.~\ref{fig:mean_std}(b), we can find that OmniVR consumes a little more time than the baseline condition in scenario 4. We think it is related to the task design in scenario 4. The task in scenario 4 is high-level and about number counting of a specific object. As a result, if the scenario is difficult to identify, the participants prefer to give very fast ``Invisible" responses. For example, P16 consumes 2.5 seconds for the ``Invisible" response with the baseline condition while consuming 16.0 seconds to finish the counting task with OmniVR.

\subsubsection{Qualitative Evaluation}
\label{subsec:qualitative}


We present participants' feedback on the reasons for their subjective ratings of VR experiences under two experimental conditions (Fig.~\ref{fig:interview}) as well as their expectations and suggestions for further improvement.

\noindent \textbf{Usability.} Generally, participants rated a higher score for the usability of the zoom-in technique in OmniVR than that in the baseline condition with a significant difference ($p < .001$). Most participants (N=7) thought that the zoom-in technique is \q{less effective in the baseline condition, but very helpful in OmniVR}, while only one participant (P15) thought that the zoom-in technique is \q{useful in both groups}. In contrast, few participants (P18) complained about \q{the more blurry effects} of zoomed-in regions than the original regions. Interestingly, P13 presented a different view for the usability of the zoom-in technique, \q{The second group of scenarios is clear enough, and I have no need to zoom in; but for the first group of scenarios, I still need to zoom in to find more details.} 

\noindent \textbf{Immersion.} Overall, participants show better satisfaction with the immersive experiences under OmniVR other than the baseline condition with a significant difference ($P < .001$).
The main reason that improves such an immersive experience is mentioned as the high-quality details recovered from OmniVR.
For example, P17 said that \q{The scenarios are natural and I feel interesting when watching these scenarios.}
However, a few participants (N=3) also expressed that there are still some cases that destroy immersion. For example, \q{the spherical distortion} (P2) makes the scene \q{unrealistic} (P3), and \q{the seriously blurry effect exists out of the focused regions} (P6). These effects are raised due to the original property of ODIs and M\"obius transformation. 
Individually, one participant (P8) complained about the scenario selection and experienced less immersion in the first scenario because it is \q{about a clock and looks like a planar scenario.}

\noindent \textbf{Cybersickness.} 
Participants reported less cybersickness with the OmniVR. Although there shows no significant difference between the two conditions, some participants (N=3) indicated that OmniVR could \q{recover more local details} and thus alleviate the \q{blurry effect}, which is the main reason to raise cybersickness.
Some other participants explained that they did not feel the obvious cybersickness difference as \q{the duration of this VR experience was not very long}.


\noindent \textbf{Workload.} Related to the lower cybersickness, participants reported lower mental and physical loads with OmniVR. In particular, OmniVR shows a significant difference ($P < .05$) compared with the baseline condition on mental cost. Some participants (N=4) mentioned the main reason as \q{the blurry effect would increase the workload simultaneously}. Specifically, two participants discussed that their mental costs would increase correspondingly \q{if the scenario is hard to identify}.



\section{Discussion}
\label{sec:discussion}

\noindent \textbf{About the zoom-in function.} In Sec.~\ref{subsec:qualitative}, we have demonstrated the effectiveness of the zoom-in function, especially in OmniVR. However, four participants (P7, P8, P12, P16) reflected that the zoom-in function is limited to fixed zoom levels. Specifically, although the zoomed-in regions could provide more local details, the objects of interest often occupy out of the FoV and are not in the center of the FoV. In this case, the participants have to adjust their viewing directions continuously to find the optimal direction. This would result in a bad immersive experience and increase the burdens in both mental and physical aspects. To improve it, we would try to learn how to select an optimal transformation by only assigning the interested objects. This might include the techniques about scene understanding techniques, such as ODI object detection.

\noindent \textbf{About the accuracy.} In four scenarios (1,2,3,5), the questions are about text recognition. There are two issues about these questions. Firstly, P12 said that ``\textit{In scenario 3, some words on the bridge are written with scrawl, making it ambiguous for recognition}". Secondly, P16 said that ``\textit{Some words might be guessed by associating them with prior knowledge}". That is, although some words are difficult to identify due to blurry effect, they might be responded rightly if the meaning of the sentence is understood by participants.

\noindent \textbf{Failure cases.} There is an interesting phenomenon in the second question of scenario 4. The question is about how many pipes are within the hands of the three standing people. As shown in the top of Fig.~\ref{fig:none_visual}, only one person (middle) takes a pipe in hand. Statistically speaking, only one participant gives the right response using OmniVR, while three participants give the right responses using the baseline condition. To further analyze, we find that two railings are along the standing people. OmniVR recovers clearer railings but fails to recover the detail of the pipe (See Fig.~\ref{fig:none_visual} middle). The railing might be mistaken as the pipe. As a result, most participants said that ``\textit{Two pipes are in the hands of standing people}". Instead, the baseline condition can not recover the detail of the pipe, and only one railing might be seen indistinctly. As a result, three participants said that ``\textit{One pipe is in the hands of standing people}". In the other question about pipes in hand from lying people (See Fig.~\ref{fig:none_visual} bottom), as OmniVR can recover the details of pipes clearly, the number of right responses in OmniVR increases obviously.

\noindent \textbf{Limitation.} Our system can recover HR and high-quality details under various user commands. However, the user commands are totally determined by the user operation, as discussed in the zoom-in function of Sec.~\ref{sec:discussion}. In addition, the streaming speed of our algorithm is also a limitation. In the current stage, we can only generate the transformed ODI in advance on GPUs according to the user commands, and then display the transformed ODI in VR.



\section{Conclusion}
\label{conclusion}

In this paper, we have presented a novel OmniVR system to enable viewers to navigate and zoom in/out effortlessly in the VR media, and have developed a learning-based algorithm to refine the visual fidelity. By conducting a comprehensive user study, our system was witnessed the following benefits: 1) Our system improved the scenario recognition for viewers by recovering the details of objects of interest; 2) Our system reduced the discomfort and helped viewers gain confidence obviously during VR navigation; 3) Our system was user-friendly in various user commands, \ie, navigation and zoom in/out. Our study revealed the importance of visual quality under various user commands in VR navigation, especially when the objects of interest were too small and required to zoom in. We release the project code of our OminVR system to inspire future studies in the community at \url{http://vlislab22.github.io/OmniVR/}. 


\bibliographystyle{IEEEtran}
\bibliography{IEEEabrv, template}

\vfill

\end{document}